%% file: main.tex
\documentclass[twocolumn]{aastex63}

\usepackage{apjfonts}
\usepackage{lineno}
\usepackage{color}
\usepackage{soul}
\usepackage{natbib}
\usepackage{hyperref}
\usepackage{amsmath}
\usepackage{xspace}
\usepackage{graphicx}
\usepackage{enumitem}
\usepackage{amssymb}
\usepackage{xifthen}
\usepackage[normalem]{ulem}
\usepackage{aas_macros}
\usepackage{amsfonts}
\usepackage{amsmath}
\usepackage{multirow}
\usepackage{dsfont}
\usepackage{hyperref}

\newcommand{\SSSSS}{${S}^5$\xspace}
\newcommand{\code}[1]{\texttt{#1}\xspace}

\hypersetup{    
  colorlinks      = {true},
  linkcolor       = {blue},
  citecolor       = {blue},
  urlcolor        = {blue},
}

\shorttitle{Indus stream}
\shortauthors{Hansen et al.}

\begin{document}
\title{${S}^5$\xspace: The destruction of a bright dwarf galaxy as revealed by the chemistry of the Indus stellar stream}

\input{authors.tex}

\correspondingauthor{T.~T.~Hansen}
\email{thansen@tamu.edu}



\begin{abstract}
The recently discovered Indus stellar stream exhibits a diverse chemical signature compared to what is found for most other streams due to the abundances of two outlier stars, Indus$\_$0  and Indus$\_$13. Indus$\_$13, exhibits an extreme enhancement in rapid neutron-capture ($r$-)process elements with $\mathrm{[Eu/Fe]} = +1.81$. It thus provides direct evidence of the accreted nature of $r$-process enhanced stars. In this paper we present a detailed chemical analysis of the neutron-capture elements in Indus$\_$13, revealing the star to be slightly actinide poor. The other outlier, Indus$\_0$, displays a globular cluster-like signature with high N, Na, and Al abundances, while the rest of the Indus stars show abundances compatible with a dwarf galaxy origin. Hence, Indus$\_0$ provides the first chemical evidence of a fully disrupted dwarf containing a globular cluster. We use the chemical signature of the Indus stars to discuss the nature of the stream progenitor which was likely a chemically evolved system, with a mass somewhere in the range from Ursa Minor to Fornax.
\end{abstract}

\keywords{Stellar abundances (1577), Dwarf galaxies (416), Milky Way stellar halo (1060),Globular star clusters (656)}

\section{Introduction \label{sec:intro}}
The detection of numerous stellar streams in large photometric surveys like the Sloan Digital Sky Survey (SDSS; \citealt{york2000,stoughton2002}) and the Dark Energy Survey (DES; \citealt{desdr1}) has made it clear that the Milky Way (MW) halo contains remnants from numerous accreted globular clusters and dwarf galaxies. Stellar streams form when satellite systems fall into the MW halo and are slowly disrupted and mixed into the MW stellar population. The streams thus provide a unique window to study the hierarchical build up of the Galaxy. With the recent addition of all sky proper motions from the $Gaia$ mission \citep{gaia2016,gaiadr2} a full chemodynamical study of the systems can be performed providing strong diagnostics of the stream progenitor systems.

In particular, the chemically distinct pattern of stars formed in globular clusters has been used to link streams to their parent object \citep{simpson2020,hansen2020a} and further explore the chemical enrichment channels of these systems \citep{casey2021}. Although detailed abundance analysis of disrupted dwarf galaxy stars is still sparse, chemical characterization of some systems has revealed these to contain stars exhibiting enhancements in rapid neutron-capture ($r$-process) elements. Specifically, \citet{aguado2020} found an overall enhancement of $\mathrm{[Eu/Fe]} \sim +0.65$ in their nine $Gaia$ Sausage/Sequoia stars analysed and \citet{gull2021} detected a range of $-0.5 < \mathrm{[Eu/Fe]} < +1.3$ in stars belonging to the Helmi debris and tails streams and the $\omega$ Centauri progenitor stream. These results suggest that the progenitors of these streams are likely relatively massive systems which experienced multiple $r$-process enrichment events such as neutron star mergers. In addition, the search for dynamical substructure in data from spectroscopic surveys such as the HK, Hamburg/ESO and LAMOST surveys, which includes a large number of stars with stellar abundances derived, have also revealed several dynamically tied groups \citep{yuan2020,limberg2021}

In this paper we investigate the chemical signature of the stellar stream Indus, which contains both an $r$-process enhanced star, Indus$\_$13, and a globular cluster like star, Indus$\_$0. The Indus stream was discovered as a stellar overdensity in DES data \citep{shipp2018} and further characterized with radial-velocity and metallicity measurements of member stars through the Southern Stellar Streams Spectroscopic Survey ($S^5$) project \citep{li2019}. Seven of the Indus stream members were selected for high-resolution follow-up and their abundances are presented in \citet{ji2020}. In this paper we present new abundances, as well as further explore those from \citet{ji2020}. Section \ref{sec:obs} describes the observations of Indus$\_$0 and Indus$\_$13. A detailed analysis of neutron-capture elements in Indus$\_$13 is presented in Section \ref{sec:param} and the results given in Section \ref{sec:results}. Section \ref{sec:discus} discusses the implications of the abundance patterns of Indus$\_$0 and Indus$\_$13 and the general chemical signature of the stream. A summary is provided in Section \ref{sec:summary}.

\begin{deluxetable}{lrrrrrrr}
\tabletypesize{\footnotesize}
\tablecaption{\label{tab:obs} Observing log for Indus$\_$0 and Indus$\_$13}
\tablehead{Obs Date & MJD & $t_{\text{exp}}$ & SNR & SNR & $v_{\text{hel}}$ &$\sigma(v)$ &$N_{\text{ord}}$\\
&  & (min) & 4500\AA & 6500\AA & (kms$^{-1}$) & (kms$^{-1}$) & }
\startdata
\multicolumn{8}{c}{Indus$\_$0}\\
\hline
2019-07-27 & 58691.16 & 50 &  25 & 47 &$ -28.8$&  0.2 & 34 \\ 
\hline
\multicolumn{8}{c}{Indus$\_$13}\\
\hline
2019-07-25 & 58689.23 & 86 &  23 & 43 &$ -58.0$&  0.3 & 35 \\ 
2020-11-17 & 59170.60 & 92 &  12 & 35 &$ -59.2$&  0.6 & 37 \\
\enddata
\end{deluxetable}

\begin{deluxetable*}{lllll}
\centerwidetable
\tablecolumns{5}
\tabletypesize{\footnotesize}
\tablecaption{\label{tab:data}Basic Data for Indus$\_$0 and Indus$\_$13}
\tablehead{Quantity & Indus$\_$0  & Indus$\_$13 & Unit & Reference}
\startdata
Gaia DR2 id & 6390575508661401216 & 6412626111276193920 & &\\
RA & 23:24:01.75 & 22:05:30.97 & hh:mm:ss.ss &\\
Dec & $-$64:02:20.8 & $-$56:30:53.4 & dd:mm:ss.s & \\
$g$ &16.20 & 17.05 & mag & \citet{desdr1}\\
$r$ & 15.60& 16.46 & mag & \citet{desdr1}\\
T$_{\rm eff}$ &5040$\pm$47 & 5063$\pm$58 & K & \cite{ji2020} \\
$\log g$  & 1.93$\pm$0.16 &2.29$\pm$0.16 & cgs & \cite{ji2020} \\
$\xi$ &1.73$\pm$0.23 & 1.59$\pm$0.16 & kms$^{-1}$ & \cite{ji2020} \\
$\mathrm{[M/H]}$ & -2.41& $-$1.91 &  & \cite{ji2020} 
\enddata
\end{deluxetable*}

\begin{deluxetable}{ccccccc}
\tablecaption{\label{tab:lines}Atomic Data and Abundances for Individual Neutron-Capture Element lines}
\tablehead{El. & $\lambda$ & $\chi$ & $\log gf$ & $\log\epsilon$ & $\sigma_{\log\epsilon}$ & ref\\
 & (\AA) & & & & (dex) &}
\startdata
\input{lines_stub}
\enddata
\tablerefs{(1) \citet{kramida2018}; (2) \citet{biemont2011}; (3) \citet{ljung2006}; (4) \citet{wickliffe1994};  (5)  \citet{kramida2018}, using HFS/IS from \citet{mcwilliam1998}; (6) \citet{lawler2001a}, using HFS from \citet{ivans2006}; (7) \citet{lawler2009};  (8)  \citet{li2007}, using HFS from \citet{sneden2009}; (9) \citet{denhartog2003}, using HFS/IS from \citet{roederer2008}; (10) \citet{lawler2006}, using HFS/IS from \citet{roederer2008}; (11) \citet{lawler2001c}, using HFS/IS from \citet{ivans2006}; (12) \citet{denhartog2006}; (13) \citet{lawler2001b}, using HFS from \citet{lawler2001d}; (14) \citet{wickliffe2000}; (15) \citet{lawler2004}, using HFS from \citet{lawler2009}; (16) \citet{lawler2008}; (17) \citet{wickliffe1997}; (18) \citet{lawler2009} for $\log gf$ values and HFS; (19) \citet{lawler2007}; (20) \citet{quinet2006}; (21) \citet{xu2007}, using HFS/IS from \citet{cowan2005}; (22) \citet{biemont2000}, using HFS/IS from \citet{roederer2012}; (23) \citet{nilsson2002}}
\tablecomments{The complete version of this Table is available online only. A short version is shown here to illustrate its form and content.
}
\end{deluxetable}

\section{Observations \label{sec:obs}}
Indus$\_$0 and Indus$\_$13 were observed in July 2019 with the Magellan/MIKE spectrograph \citep{bernstein2003} as part of the $S^5$ high-resolution follow-up program. General background on the $S^5$ project can be found in \citet{li2019}, and information on membership probability calculations and target selection for the high-resolution follow-up are described in \citet{li2019} and \citet{ji2020}. Indus$\_$13 was again observed in November 2020. All observations were executed using the 0.7" slit and 2x2 CCD binning, yielding a typical resolution of $R\sim$ 35000/28000 in the blue and red arms, respectively. The data were reduced with the CarPy MIKE pipeline \citep{kelson2003} and coadded. Table \ref{tab:obs} provides an observing log for the  observations listing the signal-to-noise ratio (SNR) per pixel, heliocentric radial velocities and associated uncertainty, and the number of orders ($N_{\text{ord}}$) used for the cross-correlation of the spectra.

\section{Stellar Parameter Determination and Abundance Analysis \label{sec:param}}
\citet{ji2020} presented abundances for $\sim$25 elements in both Indus$\_$0 and Indus$\_$13 as part of their analysis of stars across multiple stellar streams. For completeness, we list the \citet{ji2020} measurements for Indus$\_$0 including errors in Table \ref{tab:abun0}. This star exhibits an abundance pattern consistent with it having formed in a globular cluster, and we discuss the implications of this in Section \ref{sec:discus}. Indus$\_$13 exhibits significant $r$-process enhancement, such that we are able to derive abundances for 24 neutron-capture elements in addition to the measurements presented in \citet{ji2020}. These new abundances are determined following the analysis in \cite{ji2020} and using the parameters for Indus$\_$13 determined in that paper (see Table \ref{tab:data}). We provide here a brief description of the procedure and refer the reader to \cite{ji2020} for details. Effective temperatures were determined from photometry using de-reddened $g-r$ colors from the DES survey \citep{desdr1} and color-temperature relations derived from Dartmouth isochrones \citep{dotter2008}. Surface gravities were determined using the distance modulus from \citet{shipp2018} in combination with DES $g$-band magnitudes. Finally, metallicities and mictroturbulences were determined from equivalent width measurements of \ion{Fe}{2} lines in the stellar spectra. 
All abundances presented here are derived via spectral synthesis using the analysis code 
SMHR \footnote{\url{https://github.com/andycasey/smhr}} to run the 2017 version of the 1D LTE radiative transfer code \code{MOOG}\footnote{\url{https://github.com/alexji/moog17scat}}. $\alpha$-enhanced ($\mathrm{[\alpha/Fe] = +0.4}$) \code{ATLAS} model atmospheres \citep{castelli2003} were used as input, with line lists generated from \code{linemake}\footnote{\url{https://github.com/vmplacco/linemake}} updated with CH lines from \cite{masseron2014} \footnote{\url{https://nextcloud.lupm.in2p3.fr/s/r8pXijD39YLzw5T?path=\%2FCH}}. 
Solar abundances were taken from \citet{asplund2009}. Our synthesis also includes isotopic and hyperfine structure broadening, when applicable, employing the $r$-process isotope ratios from \citet{sneden2008}. Values and references for the oscillator strengths ($\log gf$) employed are listed in Table \ref{tab:lines} along with wavelengths, and the excitation potential ($\xi$) values for the lines used in the analysis. Finally, \cite{ji2020} introduced a thorough systematic and statistical abundance error analysis, which we also employ here.

\begin{figure*}
\centering
\includegraphics[scale=0.3]{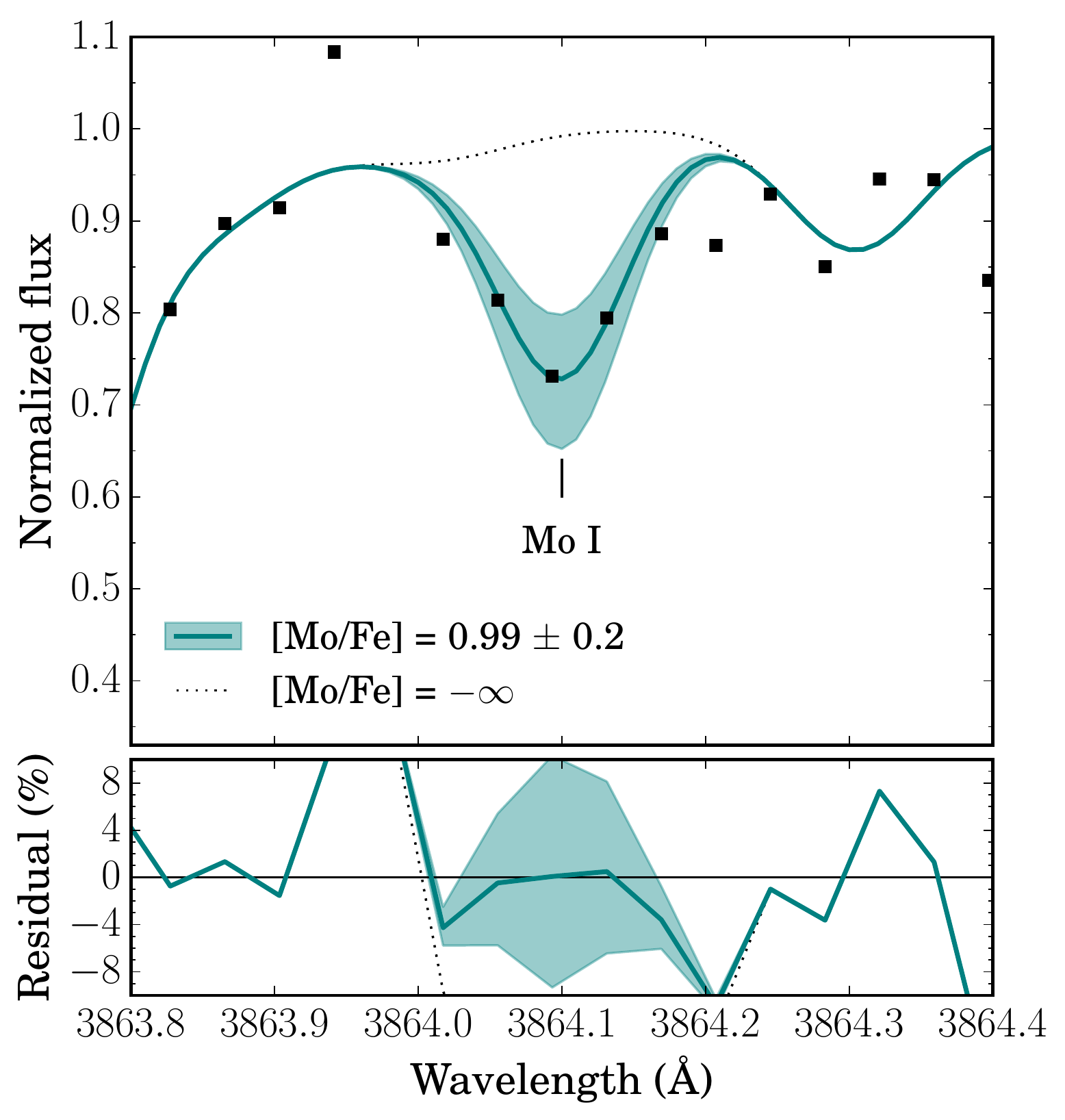}
\includegraphics[scale=0.3]{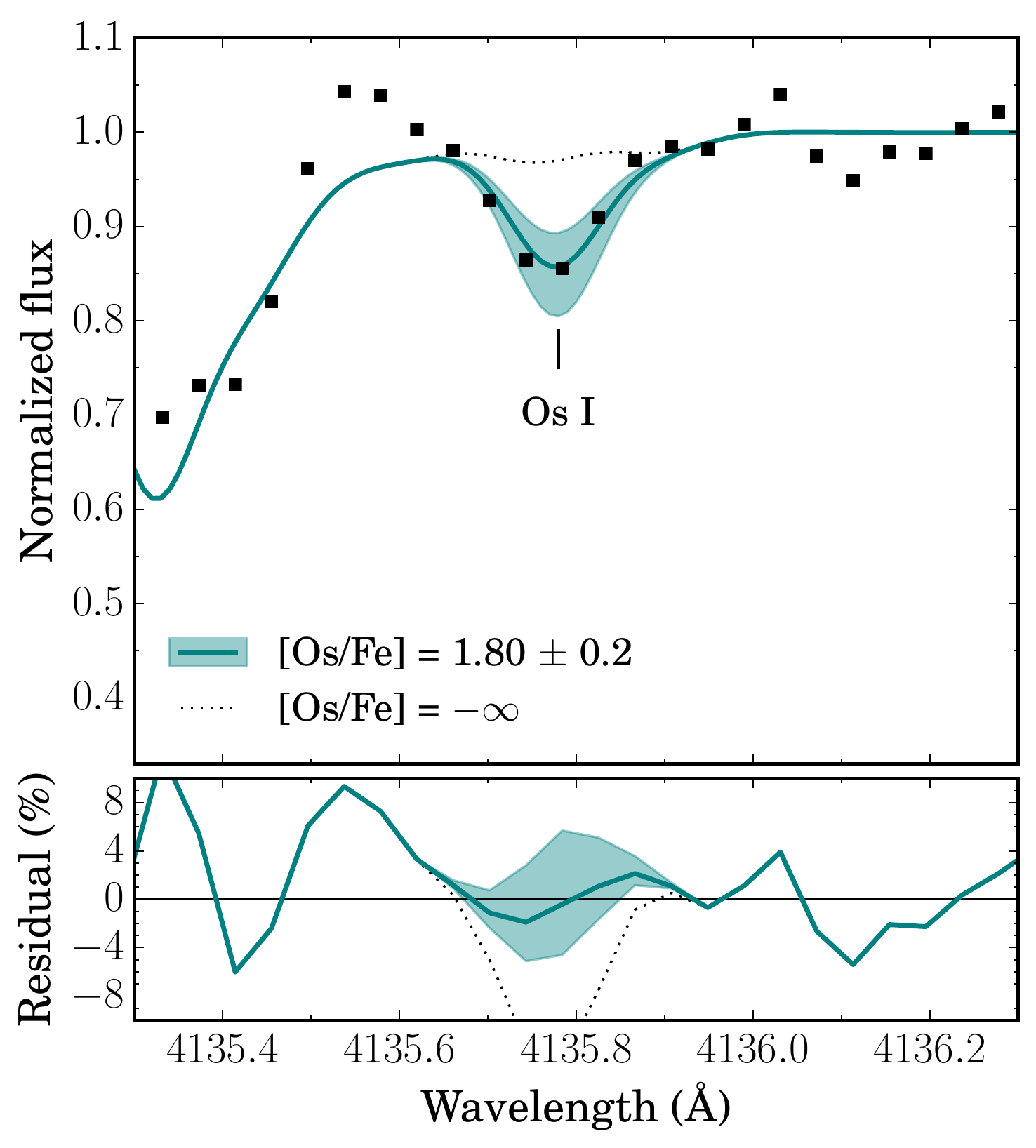}
\includegraphics[scale=0.3]{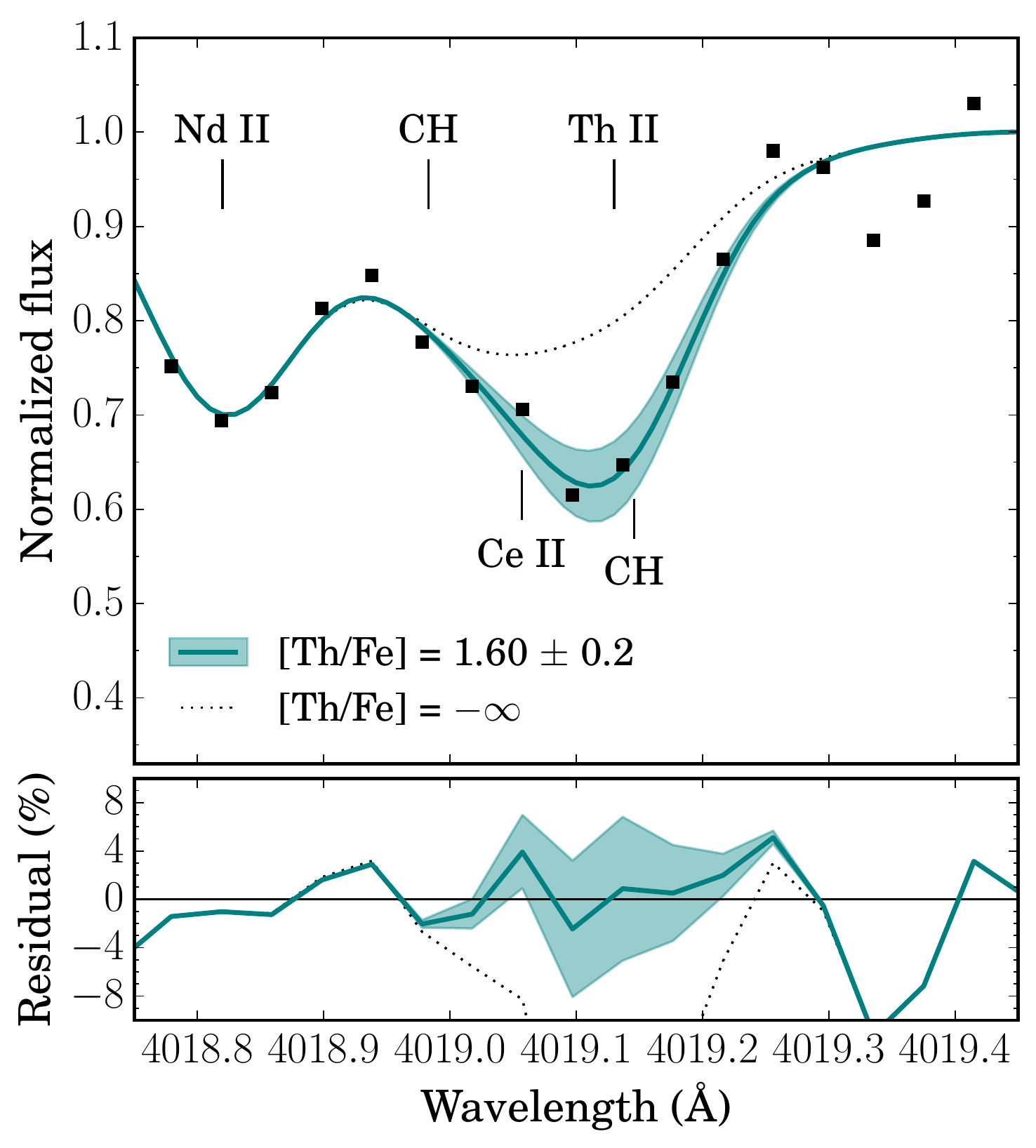}
\caption{\label{fig:synth} Synthesis of \ion{Mo}{1}, \ion{Os}{1}, and \ion{Th}{2} lines in Indus$\_$13. The top panels display the best fit synthesis (solid lines) to the observed spectrum (black squares) along with $\pm$0.2~dex uncertainty (shaded region). Residuals between the observed spectrum and the various synthesis are shown in the bottom panel. For the \ion{Th}{2} line blended features included in the synthesis are labeled.}
\end{figure*}

\begin{deluxetable*}{lrrrrrrrrrrr}
\centerwidetable
\tablecolumns{12}
\tablecaption{\label{tab:abun0}Abundance Summary for Indus$\_$0}
\tablehead{El. & N & $\log\epsilon$ & $\mathrm{[X/H]}$ & $\sigma_{\mathrm{[X/H]}}$ & $\mathrm{[X/Fe]}$ & $\sigma_{\mathrm{[X/Fe]}}$ & $\Delta_{T_{\rm eff}}$ & $\Delta_{\log g}$ & $\Delta_{\xi}$ & $\Delta_\mathrm{[M/H]}$ & $s_X$ \\
  &  &  &  & (dex) &  & (dex) & (dex) & (dex) & (dex) & (dex) & (dex)}
\startdata
\input{abun_0.txt}
\enddata
\tablecomments{Abundances are taken directly from \citet{ji2020}.}
\end{deluxetable*}

\begin{deluxetable*}{lrrrrrrrrrrr}
\centerwidetable
\tablecolumns{12}
\tablecaption{\label{tab:abun13}Abundance Summary for Indus$\_$13}
\tablehead{El. & N & $\log\epsilon$ & $\mathrm{[X/H]}$ & $\sigma_{\mathrm{[X/H]}}$ & $\mathrm{[X/Fe]}$ & $\sigma_{\mathrm{[X/Fe]}}$ & $\Delta_{T_{\rm eff}}$ & $\Delta_{\log g}$ & $\Delta_{\xi}$ & $\Delta_\mathrm{[M/H]}$ & $s_X$\\
 &  &  &  & (dex) &  & (dex) & (dex) & (dex) & (dex) & (dex) & (dex)}
\startdata
\input{abun_13.txt}
\enddata
\tablenotetext{*}{NLTE corrected: $\Delta_{\rm NLTE}$ = +0.41~dex.}
\tablecomments{Abundances C to Zn are taken directly from \citet{ji2020}}
\end{deluxetable*}

\section{Indus$\_$13 Abundance Results \label{sec:results}}
We have derived abundances for 24 neutron-capture elements, including lead (Pb) and thorium (Th), in Indus$\_$13. Synthesis of selected elements (\ion{Mo}{1}, \ion{Os}{1}, and \ion{Th}{2}) are shown in Figure \ref{fig:synth} and final abundances including errors are listed in Table \ref{tab:abun13}, where $\Delta_{\rm X}$ are the abundance errors due to stellar parameter errors and $s_X$ is the systematic error. In addition, individual line abundances are listed in Table \ref{tab:data}. For completeness we also include abundances derived for Indus$\_$13 from \citet{ji2020} in Table \ref{tab:abun13}. All abundances are derived assuming local thermodynamic equilibrium. For the majority of the neutron-capture elements, abundances are derived from the ionized states of the species, for which effects due to departures from local thermodynamic equilibrium (NLTE) are expected to be minimal. For the range of neutron-capture elements analysed here, to our knowledge, NLTE corrections only exist for Ba and Pb. For Ba, \citet{korotin2015} find average corrections of $\pm$0.1~dex for the three red \ion{Ba}{2} lines at 5853, 6141, and 6496\AA, two of which are used in this study. However, with a Ba abundance of $\mathrm{[Ba/Fe]} = +1.29$, Indus$\_$13 falls outside the grid of Ba abundances for which \citet{korotin2015} provide corrections. We therefore use the uncorrected Ba abundance. For Pb, \citet{roederer2020} derived abundances using the 2203{\AA} \ion{Pb}{2} line in the UV spectra of three metal-poor stars, yielding higher Pb abundances than results from the more commonly used \ion{Pb}{1} line at 4057{\AA} for all three stars. This confirms that \ion{Pb}{1} abundances are likely affected by NLTE effects. We therefore correct our Pb abundance using the grid from \cite{mashonkina2012}, which predicts a correction of $+0.41$~dex for the star with the best-matched parameters to Indus~13 (HD~6755; T$_{\rm eff}$ = 5100~K, $\log g$ = 2.93, and $\mathrm{[Fe/H]} = -1.68$), resulting in $\log_\epsilon$(Pb) = +1.52 for Indus$\_$13.

\begin{figure*}
\centering
\includegraphics[width=\linewidth]{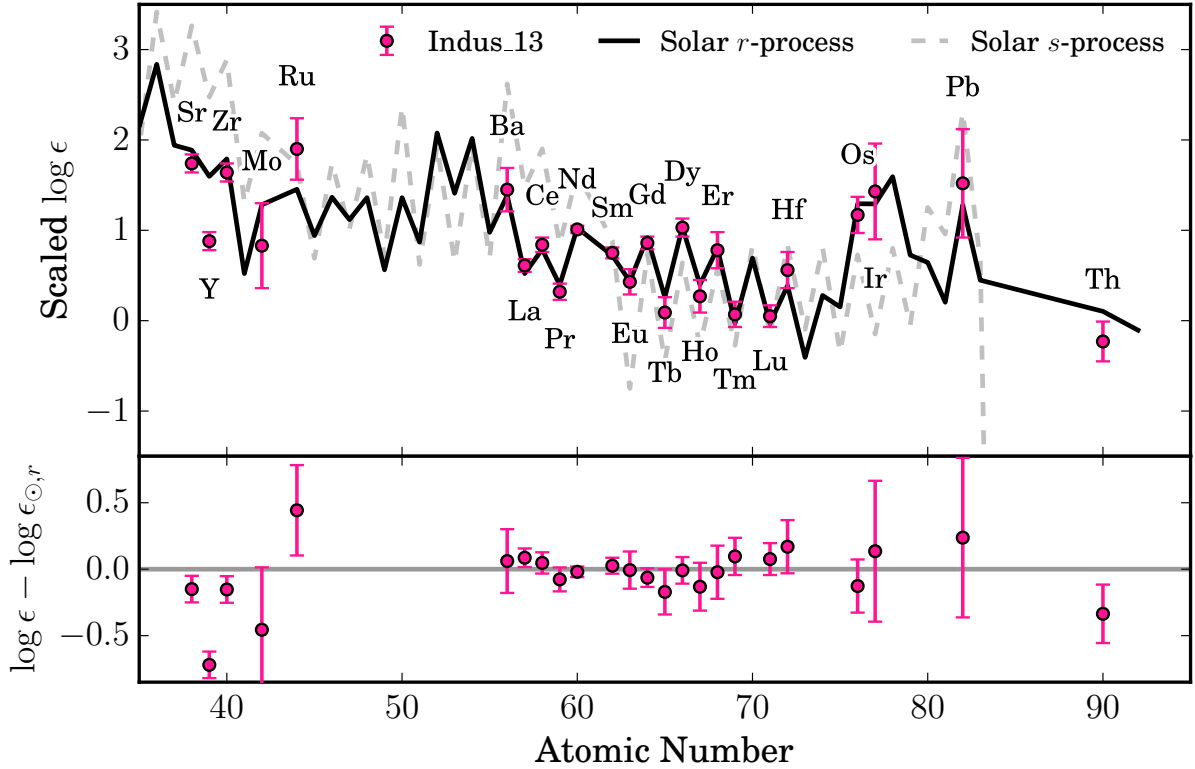}
\caption{\label{fig:rprocess} Top panel: neutron-capture element abundances derived for Indus$\_$13 (dots) along with scaled Solar-System $r$-process (black line) and $s$-process (gray line) abundance patterns \citep{sneden2008}. Bottom panel: residuals between observed abundances and scaled Solar-System $r$-process abundance pattern.}
\end{figure*}

\section{Discussion \label{sec:discus}}
\subsection{$R$-Process Signature in Indus}
\begin{figure*}
\centering
\includegraphics[width=\linewidth]{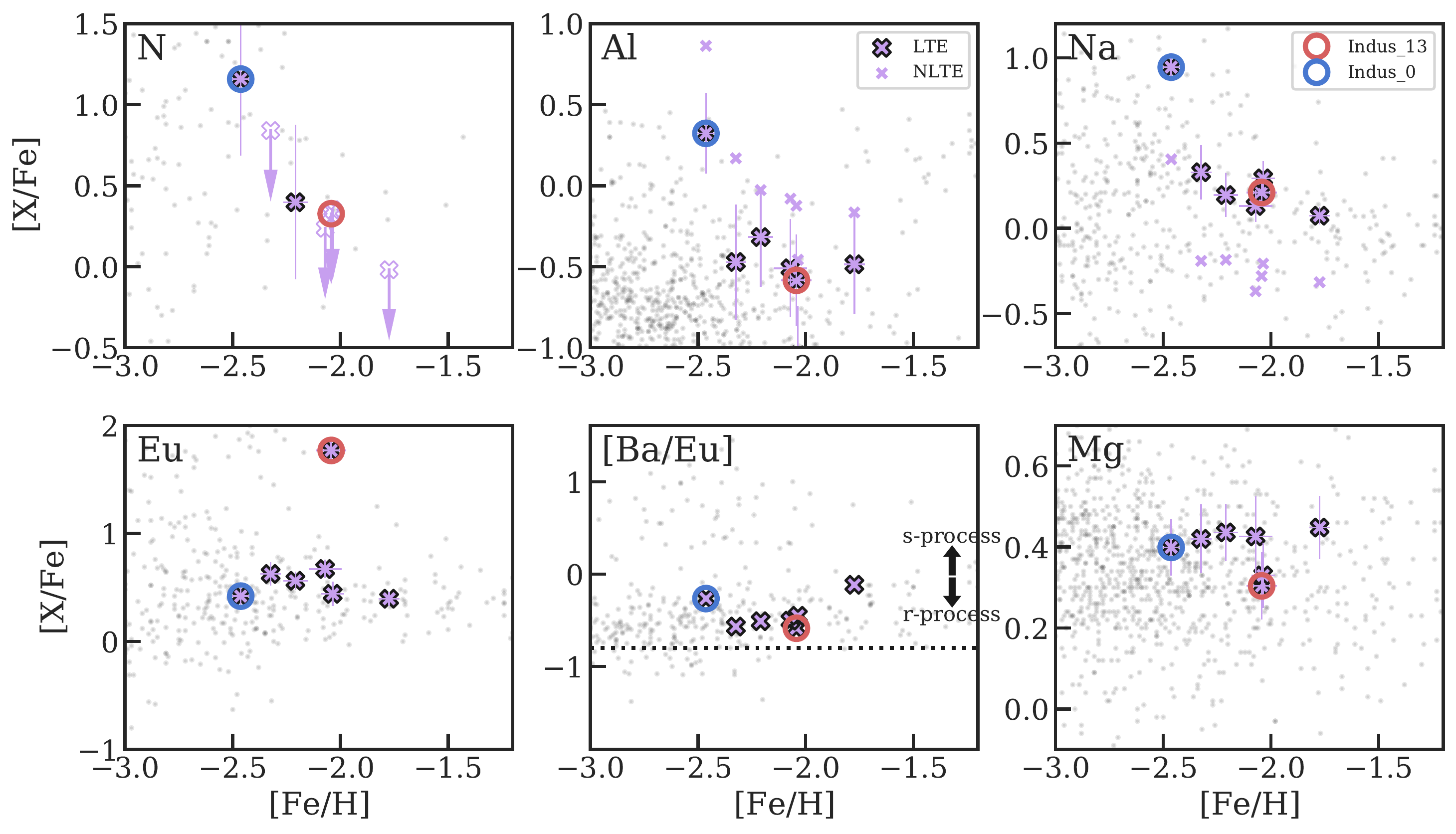}
\caption{\label{fig:abun} Light and neutron-capture element abundances for the seven Indus stars and MW halo stars from JINA base \citep{abohalima2018} highlighting the separation from the rest of the stars of Indus$\_$0 for the light elements N, Na, and Al, and of Indus$\_$13 for the neutron-capture elements Eu and Ba. NLTE corrected abundances for Na and Al are also shown in the top row displaying and general shift for all the Indus stars. Dotted line in the $\mathrm{[Ba/Eu]}$ plot (bottom middle) marks a pure $r$-process $\mathrm{[Ba/Eu]}$ ratio while the arrows mark a dominant contribution to the neutron-capture elements being from either the $s$- or $r$-process \citep{sneden2008}.}
\end{figure*}

Figure \ref{fig:rprocess} displays the derived neutron-capture element abundances for Indus$\_$13 (blue dots) along with the scaled Solar-System $r$-process (black line) and $s$-process (gray line) abundance patterns \citep{sneden2008}. An average of the abundances from La to Hf has been used for the scaling (Ba was excluded due to possible NLTE effects as described above). It is clear that for the heavy elements (La - Hf), the abundance pattern of Indus$\_$13 displays a good match to the scaled Solar-System $r$-process abundance pattern. Thus Indus$\_$13 was enriched by an $r$-process nucleosynthesis event and, with an Eu abundance of $\mathrm{[Eu/Fe]} = +1.81$, can be classified as an $r$-II star according to the classification scheme of \citet{holmbeck2020}. The other six Indus stream stars exhibit milder enhancements in Eu ($+0.40 < \mathrm{[Eu/Fe]} < +0.67$) with all having $\mathrm{[Ba/Eu]} < +0.0$ \citep[][see also Figure \ref{fig:abun}]{ji2020} consistent with enrichment from an $r$-process event. They can thus be labelled $r$-I stars. 

As can be seen from Figure \ref{fig:rprocess}, the Th abundance of Indus$\_$13 falls +0.33~dex below what is predicted from the \citet{sneden2008} Solar-System $r$-process abundance pattern. We find $\log\epsilon(\rm Th/Eu) = -0.66$, while the \citet{sneden2008} pattern predicts $\log\epsilon(\rm Th/Eu) = -0.33$. This means that Indus$\_$13 joins the sub-group of actinide-poor $r$-process enhanced stars. This group also includes the two other $r$-II stars in dwarf galaxies for which a Th abundance has been derived: COS~82 in UMi for which \citet{aoki2007} found $\log\epsilon(\rm Th/Eu) = -0.59$ and DES~J033523$-$54040 in Reticulum~II (Ret~II) for which \cite{ji2018} found the extremely low ratio $\log\epsilon(\rm Th/Eu) = -0.84$. The other extreme is the so-called actinide boost stars, where the actinides are over-produced relative to the rare-earth elements.  Examples include CS~31082$-$001 with $\log\epsilon(\rm Th/Eu) = -0.22$ \citep{hill2002} and 2MASS~J09544277$+$5246414 with $\log\epsilon(\rm Th/Eu) = -0.12$ \citep{holmbeck2018}. This abundance signature have been known to exist for some time in MW halo stars, and has also recently been detected in stars associated with the Helmi debris stream \citep{gull2021}, but actinide boost stars have yet to be discovered in any present day surviving dwarf galaxy. 

The actinide elements, Th and U, are exclusively produced via the $r$-process, and modelers have struggled to explain the variations seen in these elements in $r$-process enhanced metal-poor stars \citep{mashonkina2014}. However, recent results from $r$-process nucleosynthesis modeling suggest that the abundance patterns seen in actinide-poor and actinide-boosted stars can be produced by varying the electron fraction ($Y_e$) distribution within the same $r$-process site \citep{holmbeck2019}. The \citet{holmbeck2019} analysis was performed for the $r$-process nucleosynthesis in neutron-star mergers, which is currently the only $r$-process element production site that has been directly observed \citep{drout2017}. A similar analysis of actinide element production in other proposed $r$-process elements production sites such as magneto-rationally driven supernovae and collapsars \citep{winteler2012,siegel2019} may help to better identify the source of $r$-process elements in systems like Indus. 

Another way to investigate the source of $r$-process elements in Indus is to calculate the fraction of lanthanide elements to lighter neutron-capture elements - the so-called lanthanide fraction ($X_{\rm LA}$) - for Indus$\_$13, which can be directly compared to the lanthanide fraction found for the 2017 neutron-star merger GW170817  \citep{ji2019}. We find $\log X_{\rm LA}$= -1.33$\pm$0.2 for Indus$\_$13 (the error is dominated by systematics). This is somewhat higher than the $\log X_{\rm LA}$= -2.2$\pm0.5$ found for GW170817 but similar to the $\log X_{\rm LA}$=-1.44  found for the highly $r$-process enhanced stars in the MW halo. This discrepancy between the $\log X_{\rm LA}$ for GW170817 and the highly $r$-process enhanced MW halo stars led \citet{ji2019} to conclude that neutron-star mergers with more lanthanide rich ejecta are needed for neutron-star mergers to remain viable as the dominant $r$-process element production site. It is likely that many of the stars included in the highly $r$-process enhanced MW halo sample used for the calculation in \citet{ji2019} are accreted dwarf galaxy stars similar to Indus$\_$13. With the separation of the halo $r$-process enhanced stars into dynamically tied groups, a comparison $X_{\rm LA}$ can be done on a system-by-system basis, allowing us to better characterize the range of lanthanide fractions captured in metal-poor stars.

A further interesting feature of Indus$\_$13, in connection to its $r$-process enrichment, is its relatively high metallicity. The stars in the Indus stream span a range in  metallicity from $\mathrm{[Fe/H]} = -2.44$ to $-1.69$ \citep{ji2020}, where Indus$\_$13, with a metallicity of $\mathrm{[Fe/H]} = -1.92$, joins the metal-rich end of this range. With its relatively high metallicity, Indus$\_$13 adds to the sample of $r$-II stars in the MW with $\mathrm{[Fe/H]} > -2$ that has emerged in recent years \citep[e.g.][]{roederer2018a,hansen2018,sakari2018}. Early exploration and surveys for $r$-II stars in the MW halo found $r$-II stars only in a fairly narrow metallicity range from $\mathrm{[Fe/H]} = -3.19$ to $-2.58$, suggesting that these stars are preferentially found at the lowest metallicities.  On the other hand, the first $r$-I stars were discovered over a wider metallicity range, extending to metallicities of $\mathrm{[Fe/H]}\sim -1.5$ \citep{barklem2005}. However, the recent extensive search for $r$-process enhanced stars initiated by the $R$-Process Alliance (RPA) \citep{hansen2018} has found a number of $r$-II stars with $\mathrm{[Fe/H]} > -2$. In fact, of the 74 $r$-II stars discovered in the four RPA data release papers, 21 have $\mathrm{[Fe/H]} > -2$, corresponding to 28\%.  This percentage is very similar to that found for the 242 $r$-I stars discovered, where 70 have $\mathrm{[Fe/H]} > -2$, corresponding to 29\% \citep{hansen2018,sakari2018,ezzeddine2020,holmbeck2020}. These results strongly suggest that the metallicity distribution function is the same for $r$-I and and $r$-II stars in the halo. Thus the high metallicity of Indus$\_$13 is not unexpected or unusual since $r$-II stars, though rare, are equally likely to be found at high metallicities as $r$-I stars.

Apart from the MW, $r$-process enhanced stars have also been discovered in several classical dwarf galaxies \citep[e.g.][]{cohen2009,cohen2010}  as well as in ultra-faint dwarf (UFD) galaxies. The discovery of $r$-process enhanced stars in the UFD galaxies Ret~II \citep{ji2016,roederer2016}, Tucana~III \citep{hansen2017,marshall2019}, and Grus~II \citep{hansen2020b} supports the idea that $r$-process enhanced stars in the MW halo likely found their way into the halo through the accretion and disruption of such smaller systems \citep{brauer2019}. With the release of kinematic data from the $Gaia$ satellite, this theory has been tested: first on the smaller sample in \citet{roederer2018b}, and most recently by \citet{gudin2020} on the full literature sample that includes the large number of $r$-process enhanced stars discovered as part of the RPA. Both of these find that a number of $r$-process stars cluster in groups sharing kinematic features, suggesting that the stars were likely stripped from the same parent object. Indus$\_$13, as the first highly $r$-process enhanced star to be discovered in a disrupted dwarf galaxy stream, along with the recent discovery of $r$-process enhanced stars in the Helmi debris, Helmi tail, and $\omega$ Centauri progenitor streams \citep{gull2021}, thus provides a vital observational link supporting the accretion origin for $r$-process enhanced stars. In addition, as is pointed out by \citet{gull2021}, many $r$-process enhanced stars in the MW halo likely come from relatively massive galaxies that have experienced multiple $r$-process events and not just from UFD galaxies. As a consequence, we should expect to find stronger kinematic clustering among these stars from larger galaxies, compared with the smaller number of stars accreted from small galaxies.

\subsection{A Globular Cluster Star in Indus}

The Indus stream contains a second chemically distinct star, namely Indus$\_$0, for which \citet{ji2020} found $\mathrm{[N/Fe]} = +1.16$, $\mathrm{[Na/Fe]} = +0.94$, and $\mathrm{[Al/Fe]} = +0.32$, clearly separating it from the other Indus stars for these elements, as can be seen in Figure \ref{fig:abun}. The abundances for other elements in this star, however, are very similar to those for the rest of Indus sample. Since both Na and Al are known to be affected by NLTE effects we have also plotted the NLTE corrected abundances for Na and Al for the Indus stars in Figure \ref{fig:abun} using corrections from \citet{lind2011} and \citet{nordlander2017}, respectively. These corrections mainly result in an overall shift of the Na and Al abundances of the Indus stars and do not change the status of Indus$\_$0 as an outlier. 

With high N, Na and Al abundances, Indus$\_$0 is similar to the abundance pattern found in so-called `second generation' stars in globular clusters.  Such stars show distinct enhancements in N, Na and Al abundances coupled with depletions in C, O and Mg relative to the abundances of `first generation' or `primordial' stars in the cluster.  While the enrichment process is not understood, it is clear that nuclear-burning cycles, like the CNO, Ne-Na and Mg-Al cycles, are involved in generating the variations in the light element abundance pattern of the stars \citep[e.g.][]{gratton2019}.  The presence of these light element abundance variations is a characteristic of globular clusters, and there is no evidence of such correlated light-element abundance variations occurring in dSph or UFD field stars \citep[e.g.][]{shetrone2003,norris2017,salgado2019}. It should be noted that, with a metallicity of $\mathrm{[FeII/H]} = -2.44$, a globular cluster progenitor of Indus$\_$0 would be among the most metal-poor globular clusters known \citep{beasley2019}. However, \citet{letarte2006} have shown from high dispersion analyses of a small number of stars that similar abundance anomalies are found in the Fornax globular clusters that are only slightly more metal-rich (see discussion below). 

The spread in metallicity and the velocity dispersion of the Indus stream firmly point to it being a disrupted dwarf galaxy \citep{li2019,ji2020}. However, given that Indus$\_$0 is in the observed sample, and given the apparent lack of globular cluster second generation abundance pattern stars in the field populations of other dSph and UFD systems, the inevitable conclusion from our data is that the Indus progenitor possessed a globular cluster in which Indus$\_$0 formed.

Excluding the most massive MW satellites (Magellanic Clouds and Sagittarius) globular clusters have been detected in association with only two MW dwarf galaxies; Fornax and Eridanus 2 \citep{huang2021}. Specifically, one faint (M$_{\rm V} \approx -3.5$) cluster has been detected in the ultra faint system Eridanus 2 \citep{crnojevic2016,koposov2015}, while Fornax has six globular clusters \citep{wang2019}. The Fornax globular clusters have long been known to be old and metal-poor \citep[e.g.][]{zinn1981,buonanno1998,buonanno1999,strader2003}, and \citet{larsen2012} found that the 4 most metal-poor clusters, which have  $-2.5 < \mathrm{[Fe/H]} < -2$, match the metal-poor tail of the Fornax field star abundance distribution.  Further, \citet{larsen2014} have estimated the nitrogen abundances of red giants in these four clusters using multi-band {\it HST} imaging.  They find a range in N abundance of $\Delta\mathrm{[N/Fe]} \sim +2$~dex, with roughly equal numbers of N-normal and N-enhanced stars in each cluster.  Together with the small sample of Fornax globular cluster red giants analyzed at high dispersion discussed in \citet{letarte2006}, these results point to the existence of multiple stellar populations in the Fornax globular clusters, with properties very similar to those seen in MW globular clusters.
The inference that can be drawn from the Indus stars is then a similar picture to Fornax, with Indus$\_$0 originating in a metal-poor globular cluster in which multiple stellar populations existed, and which was associated with the Indus progenitor. 

Aside from the on-going accretion and disruption of the Sgr dwarf, which possesses at least six associated globular clusters \citep[e.g.][]{dacosta1995,law2010,massari2019,vasiliev2019}, our abundance results for the Indus stream members provide the first evidence for the accretion of a dwarf galaxy that contained at least one globular cluster. Other kinematic studies have found evidence for potentially similar events. 
For example, \citet{malhan2019} found a cocoon of stellar material associated with the globular cluster stream GD-1, which they interpret as the remnants of the dwarf galaxy with which GD-1 was accreted. Similar analysis of the Jhelum stream has revealed a two-dimensional structure for this stream with a thin dense component and a more diffuse component \citep{bonaca2019}.  These components may represent distinct but related progenitors \citep{bonaca2019}. In the context of the Indus stream results, this scenario is intriguing as it has been suggested that the Indus and Jhelum streams are two wraps of the same stream \citep{shipp2018,bonaca2019}. None of the Jhelum stars analyzed, however, display abundance patterns compatible with a globular cluster origin \citep{ji2020}.  We note for completeness that no current Galactic globular cluster has been found to have orbital characteristics similar to those of the Indus stream in the study of \citet{riley2020}.

\subsection{The Nature of the Indus Stream Progenitor}
Apart from clues from various nucleosynthesis channels, the chemical signature and specific abundance trends of a stream also provide some information on the nature of the stream progenitor.  For example, streams originating solely from globular clusters can be separated from dwarf galaxy streams using properties such as the presence or absence of iron abundance spreads, stream width and velocity dispersion \citep[e.g.\ Fig.\ 4 of][]{ji2020}.  For dwarf galaxy streams, the mean metallicity and the iron abundance of the `knee' in the ([$\alpha$/Fe], [Fe/H]) relation can also yield information on the stellar mass or luminosity of the progenitor \citep{kirby2011}.  We now use our information and that in \citet{ji2020} to place constraints on the Indus stream progenitor.

\begin{figure}[hbt!]
\centering
\includegraphics[scale=0.5]{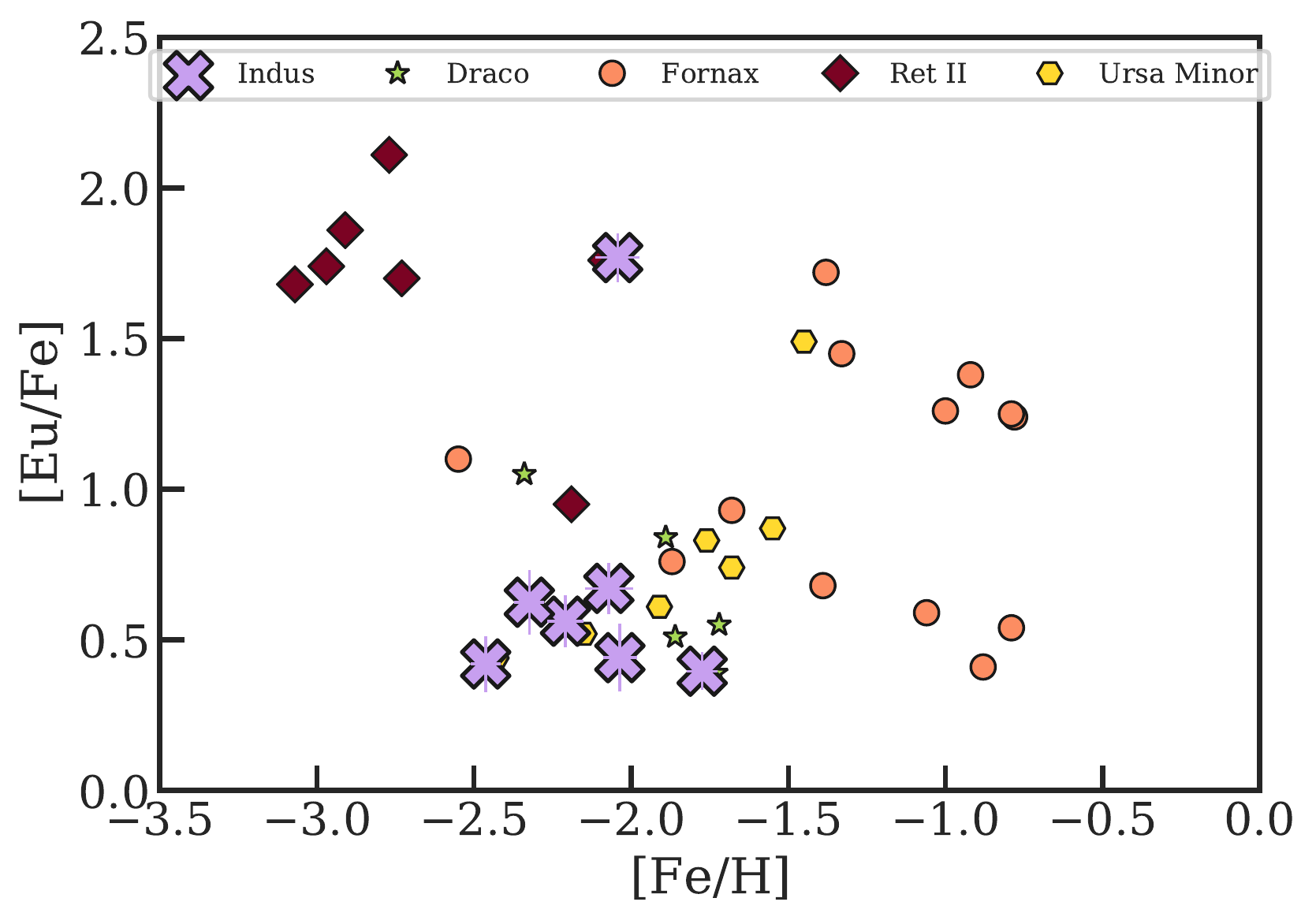}
\caption{\label{fig:EuFe} $\mathrm{[Eu/Fe]}$ as a function of metallicity for $r$-I and $r$-II stars in the Draco, UMi, and Fornax dwarf spheroidal galaxies and the ultra-faint dwarf galaxy Ret~II compared to the Indus stream stars.}
\end{figure}

Of the present-day intact dwarf spheroidal galaxies, a mixture of $r$-I and $r$-II stars has been discovered in three systems: Ursa Minor (UMi), Draco, and Fornax \citep{shetrone2001,cohen2009,letarte2010,lemasle2014}. $r$-I stars are also known to be present in Carina \citep{shetrone2003,venn2012} and Sculptor \citep{kirby2012}. Figure \ref{fig:EuFe} shows the $\mathrm{[Eu/Fe]}$ abundances as a function of metallicity for the $r$-I and $r$-II stars in Draco, UMi, Fornax, and Indus. For the dwarf galaxies, the stellar sample is that compiled in \citet{hansen2017}, where $r$-process enhanced stars in dwarf galaxies were selected to have $\mathrm{[Eu/Fe]} > +0.3$ and $\mathrm{[Ba/Eu]} < -0.4$. The latter criterion is to ensure the stars were predominantly enriched by an $r$-process event as opposed to the slow neutron-capture process ($s$-process), which also produces both Ba and Eu but in a different ratio \citep{sneden2008}. In Figure \ref{fig:EuFe} we also include three of the recently re-analysed Fornax stars from \citet{reichert2021} satisfying the above criteria. In metallicity, the Indus stream stars mostly overlap with the $r$-process enhanced stars found in Draco and UMi, whereas the $r$-I and $r$-II stars in the more massive Fornax galaxy generally have higher metallicities, and those in the small ultra-faint, Ret~II galaxy are predominantly found at lower metallicities. This suggests that the Indus's progenitor was a relatively massive galaxy with efficient chemical enrichment, similar to UMi. It should be noted, however, that the Indus stars selected for high-resolution follow-up, and thus included in this study, were preferentially selected to be metal-poor. Ongoing analysis of medium-resolution spectra for a larger sample of Indus stream stars finds a mean metallicity for the stream of $\mathrm{[Fe/H]}=-1.97$ \citep{pace2021}. Consequently, a similar analysis of a larger sample of Indus stars would likely increase the overlap with the Fornax sample.  

The chemical enrichment efficiency of a galaxy, and by proxy its mass, can also be estimated from the evolution of the $\alpha$-element abundances with metallicity. 
The galaxy will initially be enriched by a population of massive stars. These have short lifetimes that end in Type~II supernovae explosions resulting in an $\alpha$-enhanced enrichment. After this initial enrichment, lower mass and more slowly evolving stars are formed, and eventually, Type~Ia supernovae will contribute to the chemical signature of the system resulting in a decrease in the $\alpha$ to Fe ratio of the system, also known as the $\alpha$-knee \citep{tinsley1979}. The metallicity at which this transition in $\mathrm{[\alpha/Fe]}$ occurs reflects the chemical enrichment efficiency of the galaxy: in a low-mass galaxy, the $\alpha$-knee will appear at lower metallicities than in a high-mass galaxy.

All the Indus stream stars for which detailed abundances have been derived are enhanced in the $\alpha$-elements ($\mathrm{\langle[Mg/Fe]\rangle = +0.40}$, $\mathrm{\langle[Si/Fe]\rangle = +0.49}$, $\mathrm{\langle[Ca/Fe]\rangle = +0.43}$), suggesting that the location of the $\alpha$-knee of the Indus stream progenitor is at higher metallicity than the most metal-rich star included in \citet{ji2020}, which has $\mathrm{[FeII/H]} = - 1.69$. A recent abundance re-analysis of high-resolution data for stars in Fornax and UMi has found the $\alpha$-knee in these two galaxies to be located at $\mathrm{[Fe/H]} = -1.6$ and  $\mathrm{[Fe/H]} = -2.2$, respectively, and predicted the location in Draco to be $\mathrm{[Fe/H]} = -2.1$ \citep{reichert2020}. Comparing these numbers to the metallicity of the Indus stars supports the hypothesis from the Eu abundances that the Indus progenitor was likely a dwarf galaxy whose mass exceeded that of UMi but which may not have been as massive as Fornax. 

We can also use the fact that the Indus progenitor likely possessed at least one globular cluster to estimate its luminosity by calculating the normalized globular cluster specific frequency $S_{N}$ (see \citet{harris1981} and \citet{huang2021} for the definition of $S_{N}$). For a system such as Fornax with $M_{\rm V} = -13.4$ \citep{mcconnachie2012} one globular cluster results in $S_{N}$=4.5 while the same number for UMi with $M_{\rm V} = -8.8$ \citep{mcconnachie2012} is $\approx$300. An $S_{N}$ of 300 is high compared to what is generally found for the Local Group \citep{mackey2019} and seems unlikely given the results of \citet{huang2021} on the MW satellites, thus supporting our contention that the luminosity of the Indus progenitor likely exceeded that of systems such as UMi. 
This is also in agreement with the initial estimate from \cite{shipp2018} of the Indus progenitor's dynamical mass when it was stripping stars to the stream of 6.5$\times$10$^6$M$_\odot$ based on the width of the stream. In this context we also note that the postulated Indus globular cluster was unlikely to have been of low luminosity or mass.  As shown by \citet{lagioia2019} and \citet{milone2020} for example, globular clusters that possess multiple populations tend to have relatively large initial masses: only a few clusters with initial masses above $10^{5}$ M$_{\sun}$ lack multiple populations, while those that do not show multiple populations are generally of low mass \citep{lagioia2019,milone2020}.

\section{Summary} \label{sec:summary}
In this paper, we have taken a closer look at the chemical signature of the Indus stream, a disrupted dwarf galaxy recently discovered in the DES data \citep{shipp2018} and further characterised by the $S^5$ project \citep{li2019,ji2020}. In particular, a detailed neutron-capture elements abundance analysis is presented for the highly $r$-process enhanced star Indus$\_$13. The neutron-capture element abundances in this star display a good match to scaled Solar-System $r$-process abundance pattern confirming an $r$-process nucleosynthesis origin for these elements. With a slightly low abundance of the actinide element Th, Indus$\_$13 joins the sub-class of actinide-poor $r$-process enhanced stars, which also includes one of the $r$-process enhanced stars each in Ret~II and UMi \citep{ji2018,aoki2007}. The metal-rich nature Indus$\_$13 star ($\mathrm{[Fe/H]}= -1.92$) along with the general $\alpha$ enhancement seen in all the Indus stars point to the Indus progenitor being a chemically evolved dwarf galaxy. In addition to Indus$\_$13, we also discuss the globular cluster like star Indus$\_$0 which shows enhanced N, Na and Al abundances.  It is likely that this star formed in a globular cluster associated with the Indus stream progenitor. Stars with similar abundance patterns have also been seen in the globular clusters associated with the Fornax dwarf spheroidal galaxy \citep{letarte2006,larsen2014}. Taking into account all available information, we conclude that the Indus progenitor was a chemically evolved dwarf galaxy that contained at least one globular cluster and which had a mass that likely exceeded that of UMi but which was less than that of Fornax.

\acknowledgments
The authors thank the referee for a careful reading of the manuscript.

TTH acknowledges generous support from the George P. and Cynthia Woods Institute for Fundamental Physics and Astronomy at Texas A\&M University. APJ acknowledges support from a Carnegie Fellowship and the Thacher Research Award in Astronomy. ABP is supported by NSF grant AST-1813881. TSL is supported by NASA through Hubble Fellowship grant HST-HF2-51439.001 awarded by the Space Telescope Science Institute, which is operated by the Association of Universities for Research in Astronomy, Inc., for NASA, under contract NAS5-26555. A.R.C. is supported in part by the Australian Research Council through a Discovery Early Career Researcher Award (DE190100656). Parts of this research were supported by the Australian Research Council Centre of Excellence for All Sky Astrophysics in 3 Dimensions (ASTRO 3D), through project number CE170100013.

This research made extensive use of the SIMBAD database operated at CDS, Straasburg, France \cite{wenger2000}, \href{https://arxiv.org/}{arXiv.org}, and NASA's Astrophysics Data System for bibliographic information.

This work presents results from the European Space Agency (ESA) space mission Gaia. Gaia data are being processed by the Gaia Data Processing and Analysis Consortium (DPAC). Funding for the DPAC is provided by national institutions, in particular the institutions participating in the Gaia MultiLateral Agreement (MLA). The Gaia mission website is https://www.cosmos.esa.int/gaia. The Gaia archive website is https://archives.esac.esa.int/gaia.

This project used public archival data from the Dark Energy Survey (DES). Funding for the DES Projects has been provided by the U.S. Department of Energy, the U.S. National Science Foundation, the Ministry of Science and Education of Spain, the Science and Technology Facilities Council of the United Kingdom, the Higher Education Funding Council for England, the National Center for Supercomputing Applications at the University of Illinois at Urbana-Champaign, the Kavli Institute of Cosmological Physics at the University of Chicago, the Center for Cosmology and Astro-Particle Physics at the Ohio State University, the Mitchell Institute for Fundamental Physics and Astronomy at Texas A\&M University, Financiadora de Estudos e Projetos, Funda{\c c}{\~a}o Carlos Chagas Filho de Amparo {\`a} Pesquisa do Estado do Rio de Janeiro, Conselho Nacional de Desenvolvimento Cient{\'i}fico e Tecnol{\'o}gico and the Minist{\'e}rio da Ci{\^e}ncia, Tecnologia e Inova{\c c}{\~a}o, the Deutsche Forschungsgemeinschaft, and the Collaborating Institutions in the Dark Energy Survey.  The Collaborating Institutions are Argonne National Laboratory, the University of California at Santa Cruz, the University of Cambridge, Centro de Investigaciones Energ{\'e}ticas, Medioambientales y Tecnol{\'o}gicas-Madrid, the University of Chicago, University College London, the DES-Brazil Consortium, the University of Edinburgh, the Eidgen{\"o}ssische Technische Hochschule (ETH) Z{\"u}rich, Fermi National Accelerator Laboratory, the University of Illinois at Urbana-Champaign, the Institut de Ci{\`e}ncies de l'Espai (IEEC/CSIC), the Institut de F{\'i}sica d'Altes Energies, Lawrence Berkeley National Laboratory, the Ludwig-Maximilians Universit{\"a}t M{\"u}nchen and the associated Excellence Cluster Universe, the University of Michigan, the National Optical Astronomy Observatory, the University of Nottingham, The Ohio State University, the OzDES Membership Consortium, the University of Pennsylvania, the University of Portsmouth, SLAC National Accelerator Laboratory, Stanford University, the University of Sussex, and Texas A\&M University. Based in part on observations at Cerro Tololo Inter-American Observatory, National Optical Astronomy Observatory, which is operated by the Association of Universities for Research in Astronomy (AURA) under a cooperative agreement with the National Science Foundation.

\facility{Magellan:Clay}
\software{MOOG \citep{sneden1973,sobeck2011}, IRAF \citep{tody1986,tody1993}, ATLAS9 \citep{castelli2003}, linemake (https://github.com/vmplacco/linemake), NumPy \citep{numpy}, Matplotlib \citep{matplotlib}, AstroPy \citep{Astropy:13,Astropy:18}, CarPy \citep{kelson2003}, 
SMHR \citep{casey2014}}


\end{document}

%% file: authors.tex
\author[0000-0001-6154-8983]{Terese~T.~Hansen}
\affiliation{George P. and Cynthia Woods Mitchell Institute for Fundamental Physics and Astronomy, and Department of Physics and Astronomy, Texas A\&M University, College Station, TX 77843, USA}
\affiliation{Department of Physics and Astronomy, Texas A\&M University, College Station, TX 77843, USA}

\author[0000-0002-4863-8842]{Alexander~P.~Ji}
\affiliation{Observatories of the Carnegie Institution for Science, 813 Santa Barbara St., Pasadena, CA 91101, USA}

\author[0000-0001-7019-649X]{Gary~S.~Da~Costa}
\affiliation{Research School of Astronomy and Astrophysics, Australian National University, Canberra, ACT 2611, Australia}

\author[0000-0002-9110-6163]{Ting~S.~Li}
\altaffiliation{NHFP Einstein Fellow}
\affiliation{Observatories of the Carnegie Institution for Science, 813 Santa Barbara St., Pasadena, CA 91101, USA}
\affiliation{Department of Astrophysical Sciences, Princeton University, Princeton, NJ 08544, USA}

\author[0000-0003-0174-0564]{Andrew~R.~Casey}
\affiliation{Center of Excellence for Astrophysics in Three Dimensions (ASTRO-3D), Australia}
\affiliation{School of Physics \& Astronomy, Monash University, Wellington Rd, Clayton 3800, Victoria, Australia}

\author[0000-0002-6021-8760]{Andrew~B.~Pace}
\affiliation{McWilliams Center for Cosmology, Carnegie Mellon University, 5000 Forbes Ave, Pittsburgh, PA 15213, USA}

\author[0000-0001-8536-0547]{Lara~R.~Cullinane}
\affiliation{Research School of Astronomy and Astrophysics, Australian National University, Canberra, ACT 2611, Australia}

\author[0000-0002-8448-5505]{Denis~Erkal}
\affiliation{Department of Physics, University of Surrey, Guildford GU2 7XH, UK}

\author[0000-0003-2644-135X]{Sergey~E.~Koposov}
\affiliation{Institute for Astronomy, University of Edinburgh, Royal Observatory, Blackford Hill, Edinburgh EH9 3HJ, UK}
\affiliation{Institute of Astronomy, University of Cambridge, Madingley Road, Cambridge CB3 0HA, UK}
\affiliation{Kavli Institute for Cosmology, University of Cambridge, Madingley Road, Cambridge CB3 0HA, UK}

\author[0000-0003-0120-0808]{Kyler~Kuehn}
\affiliation{Lowell Observatory, 1400 W Mars Hill Rd, Flagstaff,  AZ 86001, USA}
\affiliation{Australian Astronomical Optics, Faculty of Science and Engineering, Macquarie University, Macquarie Park, NSW 2113, Australia}

\author[0000-0003-3081-9319]{Geraint~F.~Lewis}
\affiliation{Sydney Institute for Astronomy, School of Physics, A28, The University of Sydney, NSW 2006, Australia}

\author[0000-0002-6529-8093]{Dougal~Mackey}
\affiliation{Research School of Astronomy and Astrophysics, Australian National University, Canberra, ACT 2611, Australia}
\author[0000-0002-8165-2507]{Jeffrey~D.~Simpson}
\affiliation{School of Physics, UNSW, Sydney, NSW 2052, Australia}

\author[0000-0003-2497-091X]{Nora~Shipp}
\affiliation{Department of Astronomy \& Astrophysics, University of Chicago, 5640 S Ellis Avenue, Chicago, IL 60637, USA}
\affiliation{Kavli Institute for Cosmological Physics, University of Chicago, Chicago, IL 60637, USA}
\affiliation{Fermi National Accelerator Laboratory, P.O.\ Box 500, Batavia, IL 60510, USA}

\author[0000-0003-1124-8477]{Daniel~B.~Zucker}
\affiliation{Department of Physics and Astronomy, Macquarie University, Sydney, NSW 2109, Australia}
\affiliation{Macquarie University Research Centre for Astronomy, Astrophysics \& Astrophotonics, Sydney, NSW 2109, Australia}

\author[0000-0001-7516-4016]{Joss~Bland-Hawthorn}
\affiliation{Sydney Institute for Astronomy, School of Physics, A28, The University of Sydney, NSW 2006, Australia}
\affiliation{Centre of Excellence for All-Sky Astrophysics in Three Dimensions (ASTRO 3D), Australia}


\collaboration{20}{(\SSSSS Collaboration)}

%% file: lines_stub.tex
\ion{Sr}{2}& 4077.71& 0.00 & $+$0.15 & $+$1.68& 0.06& 1\\
\ion{Sr}{2}& 4161.80& 2.94 & $-$0.47 & $+$1.92& 0.18& 1\\ 
\ion{Sr}{2}& 4215.52& 0.00 & $-$0.17 & $+$1.62& 0.08& 1\\
\ion{Y}{2} & 3950.35& 0.10 & $-$0.73 & $+$1.17& 0.51& 2\\ 
\ion{Y}{2} & 4124.90& 0.41 & $-$1.38 & $+$0.71& 0.28& 2\\
\ion{Y}{2} & 4358.72& 0.10 & $-$1.15 & $+$0.85& 0.40& 2\\

%% file: abun_0.txt
C-H         &   2&   $+$6.16& $-$2.27& 0.08& $+$0.20& 0.09&   0.09&$-$0.06&   0.01&   0.04& 0.03\\ 
C-N         &   1&   $+$6.53& $-$1.30& 0.47& $+$1.16& 0.47&   0.13&$-$0.05&   0.01&   0.05& 0.00\\
\ion{O}{1}  &   1&  $<+$7.81& $-$0.88&     & $+$1.58&     &       &       &       &       &     \\
\ion{Na}{1} &   2&   $+$4.72& $-$1.52& 0.08& $+$0.94& 0.08&   0.07&$-$0.07&$-$0.07&$-$0.01& 0.06\\ 
\ion{Mg}{1} &   7&   $+$5.54& $-$2.06& 0.06& $+$0.40& 0.07&   0.04&$-$0.04&$-$0.04&   0.00& 0.09\\ 
\ion{Al}{1} &   2&   $+$4.31& $-$2.14& 0.25& $+$0.32& 0.25&   0.04&$-$0.09&$-$0.09&$-$0.06& 0.30\\ 
\ion{Si}{1} &   2&   $+$5.83& $-$1.68& 0.08& $+$0.78& 0.09&   0.07&$-$0.05&$-$0.03&   0.01& 0.00\\ 
\ion{K}{1}  &   2&   $+$3.42& $-$1.61& 0.10& $+$0.85& 0.10&   0.04&$-$0.01&$-$0.05&$-$0.01& 0.07\\ 
\ion{Ca}{1} &  15&   $+$4.28& $-$2.06& 0.06& $+$0.40& 0.07&   0.03&$-$0.00&$-$0.02&   0.00& 0.10\\ 
\ion{Sc}{2} &   8&   $+$0.82& $-$2.33& 0.12& $+$0.12& 0.12&$-$0.00&   0.05&$-$0.03&   0.01& 0.19\\ 
\ion{Ti}{1} &  17&   $+$2.73& $-$2.22& 0.08& $+$0.24& 0.08&   0.06&$-$0.01&$-$0.00&   0.01& 0.12\\ 
\ion{Ti}{2} &  31&   $+$2.79& $-$2.16& 0.07& $+$0.29& 0.09&   0.02&   0.06&   0.04&   0.02& 0.14\\ 
\ion{V}{1}  &   2&   $+$1.49& $-$2.44& 0.11& $+$0.02& 0.12&   0.02&   0.02&   0.01&$-$0.01& 0.11\\ 
\ion{V}{2}  &   2&   $+$1.72& $-$2.21& 0.10& $+$0.24& 0.11&$-$0.01&   0.06&   0.01&   0.01& 0.00\\ 
\ion{Cr}{1} &   7&   $+$3.18& $-$2.46& 0.09& $-$0.00& 0.09&   0.05&$-$0.01&$-$0.02&   0.00& 0.14\\ 
\ion{Cr}{2} &   3&   $+$3.27& $-$2.37& 0.07& $+$0.08& 0.08&$-$0.01&   0.06&$-$0.01&   0.01& 0.00\\ 
\ion{Mn}{1} &   5&   $+$2.60& $-$2.83& 0.08& $-$0.37& 0.09&   0.03&$-$0.01&$-$0.01&$-$0.00& 0.12\\ 
\ion{Fe}{1} & 118&   $+$5.04& $-$2.46& 0.05& $+$0.00& 0.00&   0.04&   0.00&   0.02&   0.01& 0.23\\
\ion{Fe}{2} &  18&   $+$5.06& $-$2.44& 0.09& $+$0.00& 0.00&   0.00&   0.06&   0.00&   0.01& 0.16\\ 
\ion{Co}{1} &   4&   $+$2.98& $-$2.01& 0.09& $+$0.46& 0.09&   0.05&   0.01&$-$0.00&   0.00& 0.04\\ 
\ion{Ni}{1} &   9&   $+$3.89& $-$2.33& 0.08& $+$0.14& 0.08&   0.05&$-$0.00&   0.00&   0.00& 0.13\\ 
\ion{Cu}{1} &   1&  $<+$2.49& $-$1.70&     & $+$0.76&     &       &       &       &       &     \\  
\ion{Zn}{1} &   2&   $+$2.22& $-$2.34& 0.11& $+$0.12& 0.11&   0.02&   0.02&$-$0.01&   0.01& 0.10\\ 
\ion{Sr}{2} &   2&   $+$0.68& $-$2.19& 0.22& $+$0.25& 0.19&   0.04&   0.03&$-$0.12&   0.02& 0.00\\ 
\ion{Y}{2}  &   4&   $-$0.30& $-$2.51& 0.09& $-$0.07& 0.10&   0.01&   0.04&$-$0.01&   0.01& 0.06\\ 
\ion{Zr}{2} &   1&   $+$0.36& $-$2.21& 0.14& $+$0.23& 0.15&$-$0.08&   0.07&$-$0.05&$-$0.05& 0.00\\ 
\ion{Ba}{2} &   5&   $-$0.11& $-$2.29& 0.16& $+$0.15& 0.14&   0.02&   0.05&$-$0.07&   0.01& 0.15\\ 
\ion{La}{2} &   3&  $<-$0.91& $-$2.01&     & $+$0.43&     &       &       &       &       &     \\
\ion{Eu}{2} &   2&   $-$1.50& $-$2.02& 0.10& $+$0.42& 0.09&   0.02&   0.07&   0.01&   0.02& 0.00\\ 
\ion{Dy}{2} &   1&   $-$0.76& $-$1.85& 0.16& $+$0.59& 0.16&   0.01&   0.06&$-$0.02&   0.02& 0.00\\ 

%% file: abun_13.txt
C-H        &   2&  $+$6.58& $-$1.85& 0.13 &$+$0.19 &0.14&   0.10&$-$0.03&   0.00&   0.12& 0.02\\ 
C-N        &   1& $<+$6.11& $-$1.72&      &$+$0.33 &    &       &       &       &       &     \\
\ion{O}{1} &   1& $<+$8.36& $-$0.33&      &$+$1.72 &    &       &       &       &       &     \\
\ion{Na}{1}&   2&  $+$4.41& $-$1.83& 0.09 &$+$0.21 &0.09&   0.08&$-$0.06&$-$0.05&   0.00& 0.04\\ 
\ion{Mg}{1}&   8&  $+$5.86& $-$1.74& 0.06 &$+$0.30 &0.08&   0.05&$-$0.04&$-$0.02&   0.01& 0.11\\ 
\ion{Al}{1}&   2&  $+$3.82& $-$2.63& 0.28 &$-$0.58 &0.28&   0.00&$-$0.03&$-$0.03&$-$0.03& 0.30\\ 
\ion{Si}{1}&   2&  $+$5.82& $-$1.70& 0.14 &$+$0.35 &0.15&   0.06&$-$0.03&$-$0.03&   0.02& 0.15\\ 
\ion{K}{1} &   2&  $+$3.75& $-$1.28& 0.12 &$+$0.76 &0.11&   0.06&$-$0.01&$-$0.06&$-$0.02& 0.06\\ 
\ion{Ca}{1}&  22&  $+$4.72& $-$1.62& 0.06 &$+$0.43 &0.08&   0.04&$-$0.00&$-$0.01&$-$0.01& 0.09\\ 
\ion{Sc}{2}&   7&  $+$1.26& $-$1.89& 0.10 &$+$0.03 &0.10&$-$0.01&   0.06&$-$0.02&   0.01& 0.10\\ 
\ion{Ti}{1}&  17&  $+$3.23& $-$1.72& 0.09 &$+$0.33 &0.09&   0.07&$-$0.01&$-$0.01&$-$0.01& 0.16\\ 
\ion{Ti}{2}&  25&  $+$3.31& $-$1.64& 0.10 &$+$0.28 &0.09&   0.02&   0.07&   0.02&   0.05& 0.15\\ 
\ion{V}{1} &   2&  $+$1.99& $-$1.94& 0.11 &$+$0.10 &0.12&   0.01&$-$0.01&$-$0.01&$-$0.07& 0.07\\ 
\ion{V}{2} &   2&  $+$2.38& $-$1.55& 0.15 &$+$0.37 &0.15&$-$0.01&   0.07&$-$0.00&   0.03& 0.00\\ 
\ion{Cr}{1}&  15&  $+$3.65& $-$1.99& 0.10 &$+$0.05 &0.09&   0.07&$-$0.01&$-$0.01&$-$0.01& 0.14\\ 
\ion{Cr}{2}&   2&  $+$3.76& $-$1.88& 0.09 &$+$0.04 &0.08&$-$0.01&   0.07&$-$0.01&   0.03& 0.00\\ 
\ion{Mn}{1}&   6&  $+$3.11& $-$2.32& 0.09 &$-$0.28 &0.10&   0.04&$-$0.02&$-$0.01&$-$0.02& 0.17\\ 
\ion{Fe}{1}& 107&  $+$5.46& $-$2.04& 0.07 &$+$0.00 &0.00&   0.04&$-$0.00&   0.00&   0.00& 0.28\\ 
\ion{Fe}{2}&  21&  $+$5.58& $-$1.92& 0.09 &$+$0.00 &0.00&$-$0.00&   0.07&   0.01&   0.04& 0.10\\ 
\ion{Co}{1}&   5&  $+$3.06& $-$1.93& 0.11 &$+$0.11 &0.11&   0.04&$-$0.00&$-$0.02&$-$0.03& 0.00\\ 
\ion{Ni}{1}&  15&  $+$4.35& $-$1.87& 0.07 &$+$0.17 &0.08&   0.05&   0.00&   0.00&$-$0.00& 0.11\\ 
\ion{Cu}{1}&   1& $<+$2.76& $-$1.43&      &$+$0.61 &    &       &       &       &       &     \\ 
\ion{Zn}{1}&   1&  $+$2.41& $-$2.15& 0.08 &$-$0.10 &0.09&   0.02&   0.03&$-$0.01&   0.02& 0.00\\ 
\ion{Sr}{2} &   3&$+$1.74&  $-$1.15& 0.10& $+$0.83& 0.12&   0.03&   0.03&$-$0.03&   0.04&0.00 \\
\ion{Y}{2}  &  17&$+$0.88&  $-$1.28& 0.10& $+$0.71& 0.13&   0.01&   0.07&$-$0.06&   0.04&0.13 \\
\ion{Zr}{2} &  13&$+$1.64&  $-$0.93& 0.10& $+$1.06& 0.12&   0.01&   0.06&$-$0.04&   0.02&0.00 \\
\ion{Mo}{1} &   1&$+$0.83&  $-$1.05& 0.47& $+$0.91& 0.47&   0.04&   0.01&   0.02&$-$0.03&0.00 \\
\ion{Ru}{1} &   2&$+$1.90&  $+$0.14& 0.34& $+$2.11& 0.34&   0.18&$-$0.12&$-$0.15&$-$0.06&0.00 \\
\ion{Ba}{2} &   3&$+$1.45&  $-$0.70& 0.24& $+$1.29& 0.19&   0.10&   0.02&$-$0.10&   0.07&0.10 \\
\ion{La}{2} &  19&$+$0.61&  $-$0.55& 0.07& $+$1.44& 0.12&   0.03&   0.06&$-$0.03&   0.04&0.00 \\
\ion{Ce}{2} &  12&$+$0.84&  $-$0.80& 0.08& $+$1.19& 0.12&   0.01&   0.06&$-$0.01&   0.03&0.09 \\
\ion{Pr}{2} &   4&$+$0.32&  $-$0.42& 0.09& $+$1.57& 0.11&   0.04&   0.07&   0.02&   0.04&0.04 \\
\ion{Nd}{2} &  41&$+$1.01&  $-$0.46& 0.04& $+$1.53& 0.13&   0.02&   0.05&$-$0.02&   0.04&0.06 \\
\ion{Sm}{2} &  39&$+$0.75&  $-$0.24& 0.06& $+$1.74& 0.12&   0.02&   0.06&$-$0.04&   0.04&0.00 \\
\ion{Eu}{2} &   8&$+$0.43&  $-$0.11& 0.14& $+$1.87& 0.13&   0.04&   0.07&$-$0.02&   0.04&0.21 \\
\ion{Gd}{2} &   8&$+$0.86&  $-$0.19& 0.07& $+$1.79& 0.13&$-$0.03&   0.08&   0.00&   0.00&0.00 \\
\ion{Tb}{2} &   2&$+$0.09&  $-$0.27& 0.17& $+$1.72& 0.15&   0.01&   0.06&   0.00&   0.04&0.00 \\
\ion{Dy}{2} &  10&$+$1.03&  $-$0.09& 0.10& $+$1.89& 0.14&   0.05&   0.07&$-$0.04&   0.04&0.00 \\
\ion{Ho}{2} &   2&$+$0.27&  $-$0.23& 0.18& $+$1.75& 0.14&   0.09&   0.07&$-$0.01&   0.06&0.00 \\
\ion{Er}{2} &   3&$+$0.78&  $-$0.14& 0.20& $+$1.85& 0.20&$-$0.04&   0.08&$-$0.07&$-$0.02&0.00 \\
\ion{Tm}{2} &   4&$+$0.07&  $-$0.01& 0.14& $+$1.98& 0.16&   0.02&   0.03&$-$0.03&   0.04&0.00 \\
\ion{Lu}{2} &   1&$+$0.05&  $-$0.05& 0.12& $+$1.93& 0.11&   0.02&   0.07&$-$0.01&   0.05&0.00 \\
\ion{Hf}{2} &   2&$+$0.56&  $-$0.29& 0.20& $+$1.69& 0.22&$-$0.08&   0.11&$-$0.01&   0.01&0.00 \\
\ion{Os}{1} &   2&$+$1.17&  $-$0.23& 0.20& $+$1.74& 0.20&$-$0.04&$-$0.15&$-$0.06&$-$0.19&0.00 \\
\ion{Ir}{1} &   1&$+$1.43&  $+$0.05& 0.53& $+$2.01& 0.54&   0.00&   0.00&$-$0.05&   0.02&0.00 \\
\ion{Pb}{1}\tablenotemark{*}&1&$+$1.52&  $-$0.65& 0.60& $+$1.32& 0.60&   0.10&$-$0.05&   0.02&$-$0.02&0.00 \\
\ion{Th}{2} &   2&$-$0.23&  $-$0.25& 0.22& $+$1.73& 0.21&   0.04&   0.03&$-$0.05&   0.02&0.00 \\